\documentclass[aps,pre,reprint,floatfix]{revtex4-2}
\usepackage[utf8]{inputenc}
\usepackage[T1]{fontenc}

\usepackage{amsmath}
\usepackage{amssymb}
\usepackage{amsfonts}
\usepackage{graphicx}
\usepackage{siunitx}

\usepackage[colorlinks=true,citecolor=blue,urlcolor=red]{hyperref}
\usepackage[normalem]{ulem}

\newcommand{\W}{\mathcal{W}}
\newcommand{\Q}{\mathcal{Q}}

\begin{document}
\title{Virtual double-well potential for an underdamped oscillator created by a feedback loop}
\author{Salamb\^{o} Dago}
\author{Jorge Pereda}
\author{Sergio Ciliberto}
\affiliation{Univ Lyon, ENS de Lyon, CNRS, Laboratoire de Physique, F-69342 Lyon, France}
\author{ Ludovic Bellon}
\email{ludovic.bellon@ens-lyon.fr}
\affiliation{Univ Lyon, ENS de Lyon, CNRS, Laboratoire de Physique, F-69342 Lyon, France}

\begin{abstract}
Virtual potentials are a very elegant, precise and flexible tool to manipulate small systems and explore fundamental questions in stochastic thermodynamics. In particular double-well potentials have applications in information processing, such as the demonstration of Landauer's principle. Nevertheless, virtual double-well potentials had never been implemented in underdamped systems. In this article, we detail how to face the experimental challenge of creating a feedback loop for an underdamped system (exploring its potential energy landscape much faster than its over-damped counterpart), in order to build a tunable virtual double-well potential. To properly describe the system behavior in the feedback trap, we express the switching time in the double-well for all barrier heights, combining for the first time  Kramer's description, valid at high barriers, with an adjusted model for lower ones. We show that a small hysteresis or delay of the feedback loop in the switches between the two wells results in a modified velocity distribution, interpreted as a cooling of the kinetic temperature of the system. We successfully address all issues to create experimentally a virtual potential that is statistically indistinguishable from a physical one, with a tunable barrier height and energy step between the two wells.
\end{abstract}

\maketitle

\section{Introduction}
Feedback traps are widely used to trap and manipulate Brownian particles in solution, and explore fundamental questions in non-equilibrium statistical mechanics of small systems~\cite{gavrilov_real-time_2014,Gavrilov_EPL_2016,PhysRevLett.94.118102,PhysRevE.86.061106}. Indeed, by controlling an external force acting on a colloidal particle as a function of its measured position, one can create a virtual potential. This is a very powerful tool, more flexible~\cite{albay_realization_2020} than its physical counterparts consisting of localized potential forces created by optical or magnetic tweezers~\cite{Berut2012,Berut2015,Hong_nano_2016,mar16,Finite_time_2020}. Feedback loops on the system's position are used in particular to study Landauer's principle, by creating double-well potentials and using the trapped particle as a memory~\cite{Bech2014,Finite_time_2020}. Within the information processing framework, lowering the dissipation seems a promising path to reduce energy costs~\cite{DagoPRL, dago2021fast, gieseler_levitated_2018, gieseler_non-equilibrium_2015}. Working with virtual potentials within underdamped dynamics thus appears as a natural endeavor. Moreover, the underdamped regime offers new insights on a wide variety of fundamental questions on the connections between feedback and thermodynamics~\cite{Seifert_2012, PhysRevLett.93.120602,PhysRevE.84.061110, rosinberg_stochastic_2015}.

Nevertheless, implementing virtual potentials in the underdamped regime is not an easy task, especially within the stochastic thermodynamics framework that requires a high measurement precision to resolve the $k_BT$ scale. Indeed, at low damping, systems are resonant and very sensitive to perturbation, noise or drift. Moreover, the feedback update delay can have strong consequences on the coupling between the system and the thermal bath~\cite{PhysRevE.86.061106, rosinberg_stochastic_2015}.

We propose in this article an electrostatic feedback designed to create virtual double-well potentials acting on a micro-cantilever, which serves as an underdamped mechanical oscillator. The system offers a flexibility and a precision never achieved before, with excellent quality in terms of position measurement and force tuning. Thanks to the thorough study of the feedback effects detailed in this article, we are able to create clean, reliable and tunable double-well potentials which outperform those produced by optical and magnetic tweezers (either physical or virtual), and have the added advantage of being analytically tractable. Therefore, this experimental work presents an unprecedented experimental tool to explore information thermodynamics, and in particular Landauer's principle in the underdamped regime.   

In the following, we detail the experimental challenges we faced to remove any bias introduced by the feedback loop. To put these challenges in context, we present a study of the response of underdamped systems to a feedback control. This study incorporates experimental and numerical simulation results, as well as a comprehensive theoretical model. The latter includes the unified and complete description of the switching time of the cantilever in the double-well potential: our expression tends towards Kramer's escape time in the high-energy barrier limit, but it also provides an adjusted model for barriers lower than the thermal energy, where Kramer's formula is no longer valid.

The article is organised as follows: we first present the experimental system and the principle of the feedback loop (section \ref{sec:principle}), before exploring the non-idealities of a real-life implementation (section \ref{sec:nonidealities}). In particular, we analyse how an hysteresis in the switches between the wells, or, equivalently, a delay in the actuation, results in a bias of the energy exchanges with the thermal bath, effectively warming or cooling the oscillator Brownian noise. From this comprehensive analysis, we define in section \ref{sec:requirements} the requirements that need to be met to mitigate imperfections. Lastly, section \ref{sec:finalsetup} describes the final implementation of the feedback loop, and shows that this loop creates a virtual potential indistinguishable from an equivalent physical one. 

\section{Virtual double-well potential: principle} \label{sec:principle}

\begin{figure}
	\centering
	\includegraphics[width=\columnwidth]{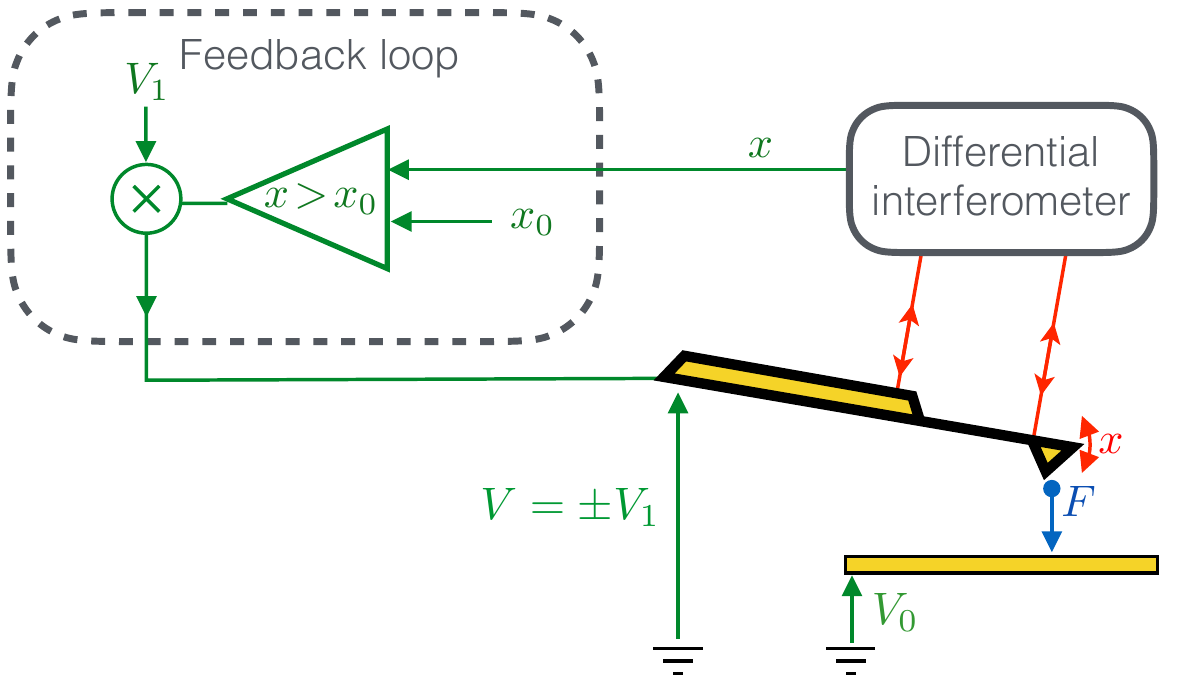}
	\caption{\textbf{Experimental system.} The conductive cantilever is sketched in yellow. Its deflection $x$ is measured with a differential interferometer~\cite{Paolino2013}, by two laser beams focused respectively on the cantilever and on its base. The cantilever at voltage $V=\pm V_1$ is facing an electrode at $V_0$. The voltage difference $V-V_0$ between them creates an attractive electrostatic force $F\propto(V-V_0)^2$. The dashed box encloses the feedback controller, consisting of a comparator and a multiplier, which create the double-well potential.}
	\label{schema_bloc}
\end{figure}

\begin{figure}
	\centering
	\includegraphics[width=\columnwidth]{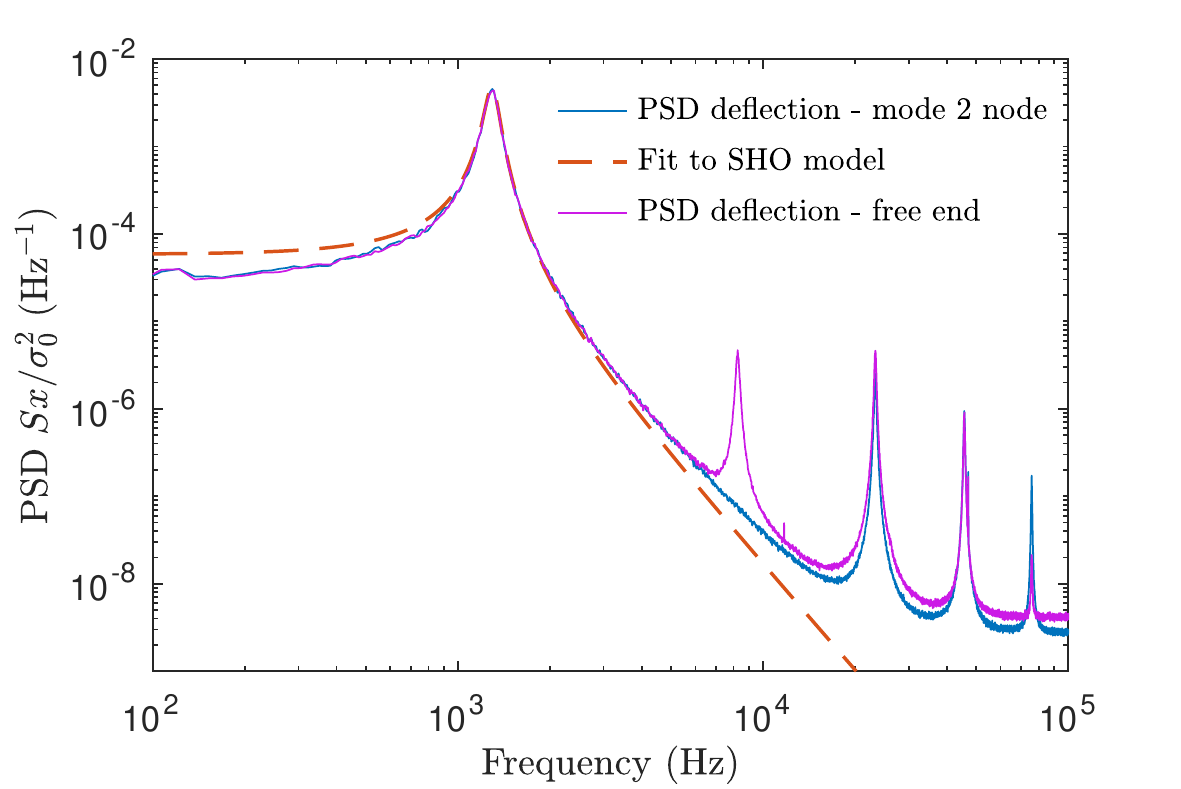}
	\caption{\textbf{Power Spectral Density (PSD) of the cantilever deflection in a single well.} Measured PSD of the thermal noise driven deflection with no feedback ($V_1=0$, solid lines), and best fit by the theoretical spectrum of a Simple Harmonic Oscillator (SHO, dashed line). The second deflection mode,  visible at $\SI{8}{kHz}$ when the laser beam is focused at the free end of the cantilever (magenta), is successfully hidden by focusing the laser beam on the node of this mode (blue). At frequencies up to $\SI{10}{kHz}$, the cantilever behaves like a SHO at $f_0=\SI{1270}{Hz}$, with a quality factor $Q=10$. We infer from this measurement the variance $\sigma_0^2 = \langle x^2 \rangle = k_B T_0 / k$, used to normalize all measured quantities.}
	\label{FigPSD}
\end{figure}

As sketched in Fig.~\ref{schema_bloc}, the underdamped oscillator is a conductive cantilever~\cite{OCTO1000S} mounted in a closed airtight chamber at room temperature $T_0$. The chamber minimises all air flows induced drifts in the measurement, and can also used as a vacuum chamber to modulate the pressure, thus the resonator quality factor, at will. The cantilever deflection $x$ is measured with very high accuracy and signal-to-noise ratio by a differential interferometer~\cite{Paolino2013}. The Power Spectral Density (PSD) of the thermal fluctuations of $x$ is plotted in Fig.~\ref{FigPSD}: the fundamental mode dominates by 3 orders of magnitude the higher-order deflection modes of the cantilever. The second deflection mode at $\SI{8}{kHz}$ is conveniently removed from the measured signal by focusing the sensing laser beam on its node, at around $0.78\%$ of the cantilever length. This simple adjustment helps in having a physical system very close to an ideal Simple Harmonic Oscillator (SHO). The fit of this PSD with the theoretical thermal noise spectrum of a SHO leads to its resonance frequency $f_0=\omega_0/2\pi=\SI{1270}{Hz}$ and quality factor $Q=m\omega_0/\gamma=10$, where $m$, $k=m \omega_0^2$ and $\gamma$ are respectively: the mass, stiffness and damping coefficient of the SHO. The slight difference between the measurement and the model is due to frequency dependency of the viscous damping of the cantilever in air~\cite{Bellon-2008}. From the PSD we compute the variance at equilibrium $\sigma_0^2 = \langle x^2 \rangle = k_B T_0 / k \sim \SI{1}{nm^2}$, which is used as length scale.

Two time scales typically describe an underdamped system: its natural oscillation period $\mathcal{T}_0=f_0^{-1}\sim\SI{0.8}{ms}$ (comparing the inertial and elastic terms), and its relaxation time $\tau_r=2Q/\omega_0\sim\SI{2.5}{ms}$ (comparing the inertial and damping terms). We add a third one, the time scale of position relaxation~\cite{ChupeauUD}, which compares the damping and elastic terms: $\tau_\gamma=\gamma/k=1/(Q\omega_0)\sim\SI{13}{\mu s}$. Due to its oscillating nature, the resonator explores the potential energy landscape typically every $\mathcal{T}_0$, and the dissipative part can be sensitive to changes in energy down to $\tau_\gamma$. This position relaxation $\tau_\gamma$ is much faster than that of most over-damped systems used to create double wells in stochastic thermodynamics, namely colloidal particles optically trapped in water. Since inertia can be neglected in these systems, their response time is set by $\tau_\gamma$, and typically amounts to $\SI{30}{ms}$~\footnote{$\tau_\gamma$ is proportional to the power of the laser that creates the trap and can vary widely, ranging from $\sim\SI{0.1}{ms}$ to $\SI{100}{ms}$ in the many optical tweezer setups that have been used to study stochastic thermodynamics. We retain $\SI{30}{ms}$ as the typical time scale of experiments tackling Landauer's bound~\cite{Berut2012, Berut2015, BerutEPL2013, Finite_time_2020, Gavrilov_EPL_2016, gavrilov_real-time_2014, PhysRevE.86.061106}}.

In order to use the cantilever as a one-bit memory, we need to confine its motion in an energy potential consisting of two wells separated by a barrier, whose shape can be tuned at will. This potential $U$ is created by a feedback loop, which compares the cantilever deflection $x$ to an adjustable threshold $x_0$. After having multiplied the output of the comparator by an adjustable voltage $V_1$, the result is a feedback signal $V$ which is $+V_1$ if $x>x_0$ and $-V_1$ if $x<x_0$. The voltage $V$ is applied to the cantilever which is at a distance $d$ from an electrode kept at a voltage $V_0$. The cantilever-electrode voltage difference $V_0\pm V_1$ creates an electrostatic attractive force $F=\frac{1}{2}\partial_dC(d)(V_0 \pm V_1)^2$~\cite{Butt-2005}, where $C(d)$ is the cantilever-electrode capacitance. Since $d\gg \sigma_0$, $\partial_dC(d)$ can be assumed constant. We apply $V_0\sim \SI{100}{V}$ and $V_1\ll V_0$ so that, to a good approximation, $F\propto \pm V_1$ up to a static term. This feedback loop results in the application of an external force whose sign depends on whether the cantilever is above or below the threshold $x_0$. As long as the reaction time $\tau_d$ of the feedback loop is very fast (at most a few $\SI{}{\mu s}$), the switching transient is negligible: $\tau_d \ll \tau_\gamma, \mathcal{T}_0, \tau_r$. As a consequence, the oscillator evolves in a virtual static double-well potential, whose features are controlled by the two parameters $x_0$ and $V_1$. Specifically, the barrier position is set by $x_0$ and its height is controlled indirectly by $V_1$, which sets the wells centers $\pm x_1 = \pm V_1 \partial_dC(d)V_0/ k$. The potential energy constructed by this feedback is:
\begin{align}
U(x,x_0,x_1)=\frac{1}{2} kx \big(x-S(x-x_0)x_1\big)^2,
\label{eq_U(z,z0,z1)}
\end{align}
where $S$ is the sign function: $S(x)=-1$ if $x<0$ and $S(x)=1$ if $x>0$. In the following, unless we specify otherwise, we will always consider the case of a symmetric potential, corresponding to $x_0=0$.

\begin{figure}
	\centering
	\includegraphics[width=\columnwidth]{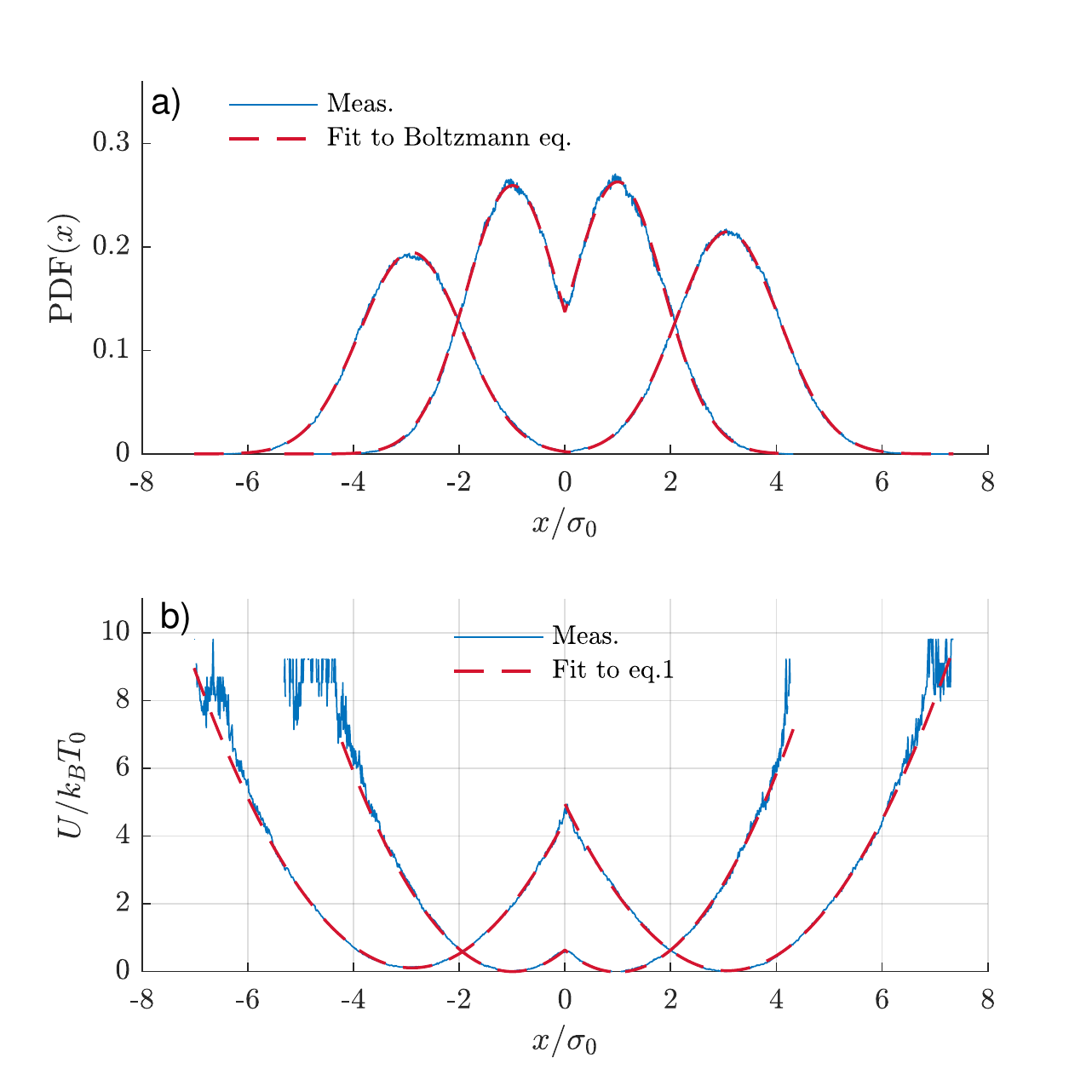}
	\caption{\textbf{(a) Probability Density Function (PDF) of $x$.} The PDF of $x$ (blue) measured during a $\SI{10}{s}$ acquisition with the feedback on, with $x_0=0$ and two values of $V_1$ adjusted to have respectively a $5\,k_BT_0$, and a $0.5\,k_BT_0$ energy barrier height. The fit using the Boltzmann equilibrium distribution with the potential shape in Eq.~\eqref{eq_U(z,z0,z1)} (dashed red) is excellent. \textbf{(b) Double-well potential energy.} The measured potentials (blue) are inferred from the PDF of $x$ in a) and the Boltzmann distribution. The high noise level for large values of $U$ stems from the bad sampling of regions with low probability. We obtain as expected the $5\,k_BT_0$ and $0.5\,k_BT_0$ barriers corresponding to the two values of $V_1$. The fits using Eq.~\eqref{eq_U(z,z0,z1)} are again excellent (dashed red).}
	\label{FigDoubleWell}
\end{figure}

The two degree of freedom of the underdamped system, the deflection $x$ and the velocity $v=\dot x$, are considered as random variables of a stochastic process. They are ruled by a Langevin equation (Eq.~\eqref{langevingen} in appendix \ref{AppendixB}), or equivalently characterized by the Probability Density Function (PDF) $P(x,v,t)$ for finding the cantilever in position $x$, and velocity $v$ at time $t$, whose dynamics is given by Kramer's equation~\cite{KRAMERS}:
\begin{align}
\partial_tP+v\partial_xP-\frac{\partial_xU}{m}\partial_vP=\frac{\gamma}{m}\partial_v(vP)+k_BT_0\frac{\gamma}{m^2}\partial^2_vP,
\label{Kramer}
 \end{align}
As the potential $U(x,x_0,x_1)$ does not depend on the speed, the equilibrium PDF of the velocity in the double-well is the same as the one in a single harmonic well and scales as a Gaussian of variance $k_BT_0/m$: $\mathrm{PDF}(v)\propto e^{-mv^2/(2k_BT_0)}$. The stationary Boltzmann distribution therefore factorizes the equilibrium $x$ and $v$ PDF:
 \begin{align}
 P_{eq}(x,v) &=\textrm{PDF}(x)\times \textrm{PDF}(v) \\
 &\propto e^{-\frac{U(x)}{k_BT_0}} \times e^{-\frac{mv^2}{2k_BT_0}}
 \end{align}
 
The potential in Eq.~\eqref{eq_U(z,z0,z1)} can be experimentally measured from the PDF of $x$ and the Boltzmann equilibrium distribution: $U(x)=U_0-k_BT_0 \ln [\mathrm{PDF}(x)]$, with $U_0$ an arbitrary constant. Fig.~\ref{FigDoubleWell} presents two examples of an experimental symmetric double-well potential generated by the feedback loop, tuned to have a barrier of $\frac{1}{2}kx_1^2 = 5k_BT_0$ and $0.5k_BT_0$ (respectively $x_1=\sqrt{10}\sigma_0$ and $x_1=\sigma_0$). The dashed red line is the best fit with Eq.~\eqref{eq_U(z,z0,z1)}, demonstrating that the feedback-generated potential behaves as a static one, in terms of the position PDF.

The experimental challenge undertaken in this work is to build a proper virtual potential identical to a physical one: the feedback loop should have no noticeable effect on the position and velocity equilibrium distributions.

\section{Virtual double-well potential: practical non-idealities} \label{sec:nonidealities}

\begin{table}[!ht]

\begin{tabular}{ |c | c | c | c | c | r |}
\hline
   Setup & Comparator & Filter  & Defect & Main bias\\
   &  &  bandpass &  & \\
   \hline
   1 & TS3022 & $\SI{1}{MHz}$ & Hysteresis:&Cooling\\
    & & & $h=0.15 \sigma_0$& \\
       \hline

   2 & LM219 & No Filter & Early trigger: &Warming\\
    & & & $h<0$& \\
       \hline

   Final & LM219 & $\SI{1}{MHz}$ &$h\sim0$ & No bias\\ 
   \hline
 \end{tabular}
 \caption{Setup 1, setup 2, and final setup distinctive features.}
 \label{Tab_setup}

\end{table}

An ideal feedback loop comparator satisfies three requirements: it presents no measurement noise, it is immediate, and it always switches exactly at the prescribed $x_0$ position. In real comparators, however, those three requirements compete with each other, and a tradeoff between them needs to be found. For example, a high-frequency measurement noise causes the comparator to switch at inexact positions. It is therefore common to low pass filter the input signal to remove this noise, at the expense of introducing a delay in the switching time. Alternatively, one can reduce the effect of noise by introducing an artificial hysteresis around the threshold, larger than the noise amplitude, but in this case the switching between wells doesn't occur at the appropriate position. In the next subsections, we study the consequences of each of these non-idealities.

\subsection{Hysteresis}

\subsubsection{Experimental observation}

One major experimental challenge lies in the comparator hysteresis. To illustrate its consequences, we use the setup 1 whose circuit is detailed in section~\ref{sec:finalsetup} (see Tab.~\ref{Tab_setup} and Fig.~\ref{electric-setup}). In this case we measure an average hysteresis of $h=0.15 \sigma_0$: the voltage switches upward from $-V_1$ to $V_1$ when the position crosses $x_0+h$ from below, and downward when crossing $x_0-h$ from above. This hysteresis is likely due to the use of the comparator outside its nominal regime in terms of voltage ranges. 

 \begin{figure}[!htb]
 \begin{center}
 \includegraphics[width=\columnwidth]{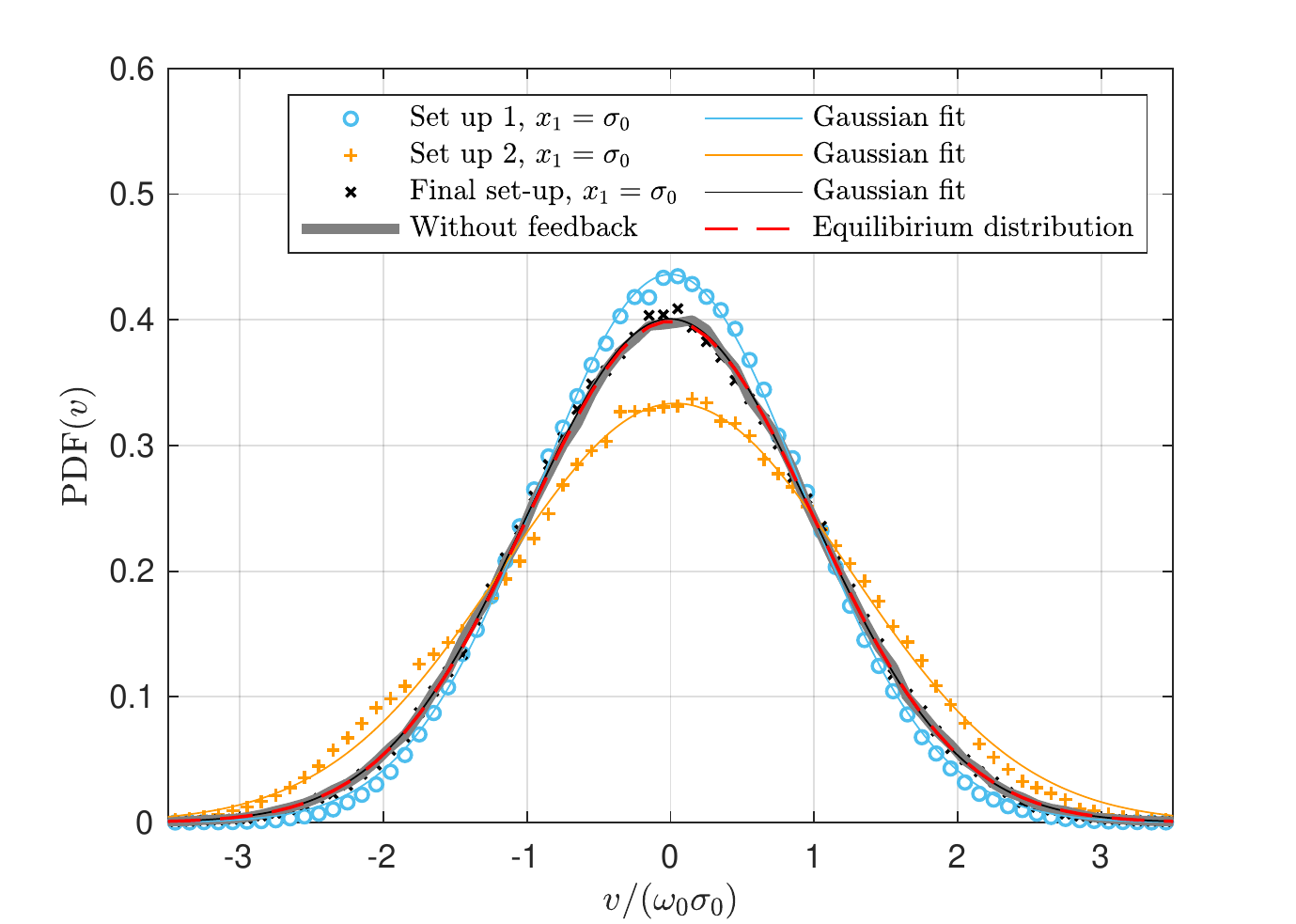}
 \end{center}
 \caption{\textbf{PDF of the oscillator speed.} Experimental PDF of $v$ for $x_1=\sigma_0$ inferred from a $\SI{10}{s}$ acquisition using setup 1, setup 2 and the final setup (see Tab.~\ref{Tab_setup}), respectively in blue, orange and black markers. Each time, the best Gaussian fit is superimposed in plain line: the fit is excellent for the setups with positive or zero hysteresis (setup 1 and final setup). Regarding setup 2 (negative hysteresis), the Gaussian fit is not as good, but remains satisfactory, and the higher moments of the experimental PDF are close to the Gaussian vanishing values: respectively $-0.08$ and $-0.3$ for the skewness and the excess kurtosis. Finally, we superimpose in grey thick line the experimental PDF without feedback, which perfectly matches the equilibrium distribution (dashed red line). It is worth noticing that the final setup (black) also ideally reproduces the equilibrium distribution.} 
 \label{FigPDFspeed}
\end{figure} 

A comparator hysteresis has an effect on the velocity distribution of the system, as illustrated in Fig.~\ref{FigPDFspeed}. While the speed PDF keeps a satisfactorily Gaussian shape for the different setups, its variances are altered compared to the equilibrium distribution perfectly matched without feedback. Therefore, the velocity variance turns out to be an adequate observable to summarize the effect of the hysteresis on the velocity distribution. The hysteresis should also alter the PDF of position for nearby wells, in particular around the threshold cusp (rounding effect), but it is a tiny effect, hard to observe experimentally.

Let us introduce the kinetic temperature $T$ of the system defined through the velocity variance: $\sigma_v^2=\langle v^2 \rangle =k_B T/m$. At equilibrium in a bi-quadratic potential, the kinetic temperature should match the bath temperature $T_0$ as prescribed by the Boltzmann distribution. To facilitate the reading we introduce the ratio $\theta=T/T_0$, so that the velocity standard deviation simplifies into $\sigma_v=\sqrt{\theta}\omega_0 \sigma_0$. 

\begin{figure}[!htb]
 \begin{center}
 \includegraphics[width=\columnwidth]{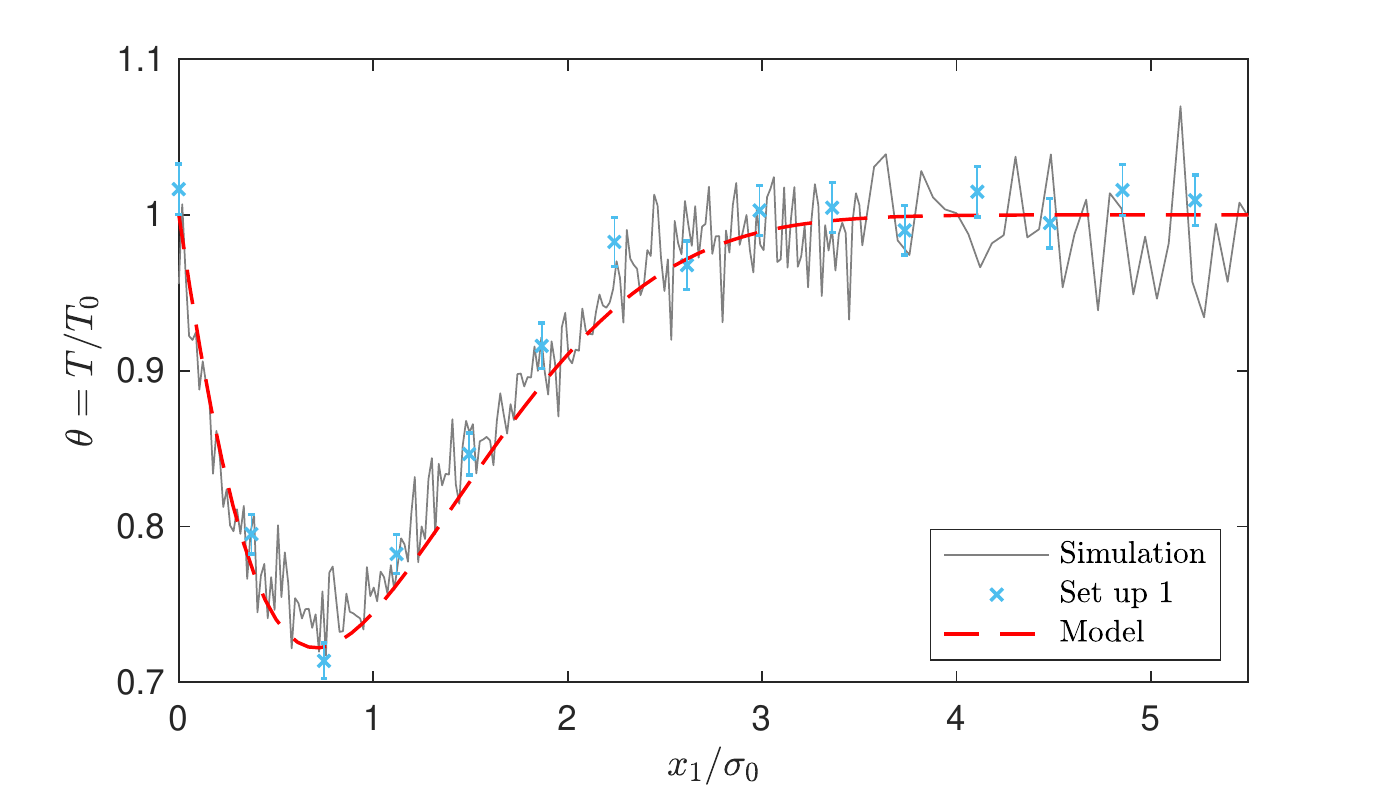}
 \end{center}
 \caption{ \textbf{Kinetic temperature with hysteresis (setup 1).} The ratio $\theta=T/T_0$ is plotted as a function of the distance between the wells $x_1$. Blue markers: experimental data obtained from setup 1 with a typical hysteresis $h=0.15 \sigma_0$ at each switch. Grey line: simulation data from $N_{\textrm{sim}}=200$ iterations of $30/f_0$ long trajectories of the cantilever evolving in a potential created with a $h=0.15 \sigma_0$ hysteresis on the threshold. Dashed red line: the numerical solution of Eq.~\eqref{eqhyst} perfectly predicts the hysteresis consequences on the temperature.} 
 \label{theta_2_models}
\end{figure}

We measure the kinetic temperature evolution through the velocity variance for different distances between the wells. The experimental results plotted on Fig.~\ref{theta_2_models} show a cooling  of the system when the wells are close. We propose in the next paragraphs a theoretical model that supports this observation.

\subsubsection{Theoretical model}

We model the consequences on the system temperature using the infinitesimal energy balance equation, with $K=\frac{1}{2} m v^2$ the kinetic energy, $\W$ the stochastic work and $\Q$ the stochastic heat~\cite{sek10,Seifert_2012,sek66,DagoPRL}: 
\begin{align}
\frac{dU}{dt}+\frac{dK}{dt}&=\frac{d\W}{dt}-\frac{d\Q}{dt} \label{eqbalance}
\end{align}
This energy balance is the starting point of the model developed in this article to link the feedback hysteresis to the system temperature, similarly to the approach followed in the theoretical description of feedback cooling~\cite{gieseler_levitated_2018,gieseler_non-equilibrium_2015}.

In a stationary state when no external work is performed ($\langle \W \rangle =0$) there is no kinetic energy evolution on average ($\langle dK/dt\rangle=0$), so that using the heat expression~\eqref{Qstat} derived in Appendix~\ref{AppendixB}, Eq.~\eqref{eqbalance} reduces to:
\begin{align}
\langle \frac{dU}{dt} \rangle &=\frac{\omega_0}{Q}k_B T_0 (1-\theta) \label{eqbalancesimp}
\end{align}

If there is a switching hysteresis, the comparator triggers only when $x=\pm h$ (sign depending on origin) instead of $x=x_0=0$. The cantilever overreaches the barrier at each crossing. This implies an extra distance travelled by the cantilever in the initial well (centred on $\pm x_1$) before the feedback makes it switch in the second well. This extra distance corresponds to a potential energy step: 
\begin{align}
\Delta U_h &=\frac{1}{2}k\big[(h+x_1)^2-(h-x_1)^2\big] \nonumber \\
&=2kx_1h
  \label{DeltaU}
\end{align}
This amount of potential energy is lost each time the cantilever crosses the barrier. Between the crossings, the system thermalizes in contact with the heat bath. Thus the system is always out-of-equilibrium and reaches a steady state characterized by the kinetic temperature $T$. The latter allows the warming heat influx from the thermostat to compensate on average the energetic losses caused by the hysteresis at each barrier crossing. It only remains to express the average heat flux corresponding to these discrete energetic losses: we need to quantify how often on average the cantilever crosses the threshold. In appendix \ref{AppendixA}, we derive the crossing rate $\Gamma$ of the potential barrier $\mathcal{B}$ for a system at temperature $T$: 
\begin{equation} \label{Gammamodel}
\Gamma ( \mathcal{B},T) = \omega_0 \frac{\mathcal{B}}{k_B T} \int_1^\infty \frac{\exp(-\epsilon\frac{\mathcal{B}}{k_BT})}{\pi + 2 \sin^{-1}(\epsilon^{-1/2})} d\epsilon
\end{equation} 

Using Eq.~\eqref{Gammamodel} applied to the barrier energy $\mathcal B =\frac{1}{2}k(x_1+h)^2$, we can express the potential contribution in Eq.~\eqref{eqbalancesimp} and derive:
\begin{align}
\Gamma(\frac{1}{2}k(x_1+h)^2,\theta T_0) \times \Delta U_h &=\frac{\omega_0}{Q}k_B T_0 (1-\theta) \label{eqhyst}
\end{align}

\begin{figure}[!htb]
 \begin{center}
 \includegraphics[width=\columnwidth]{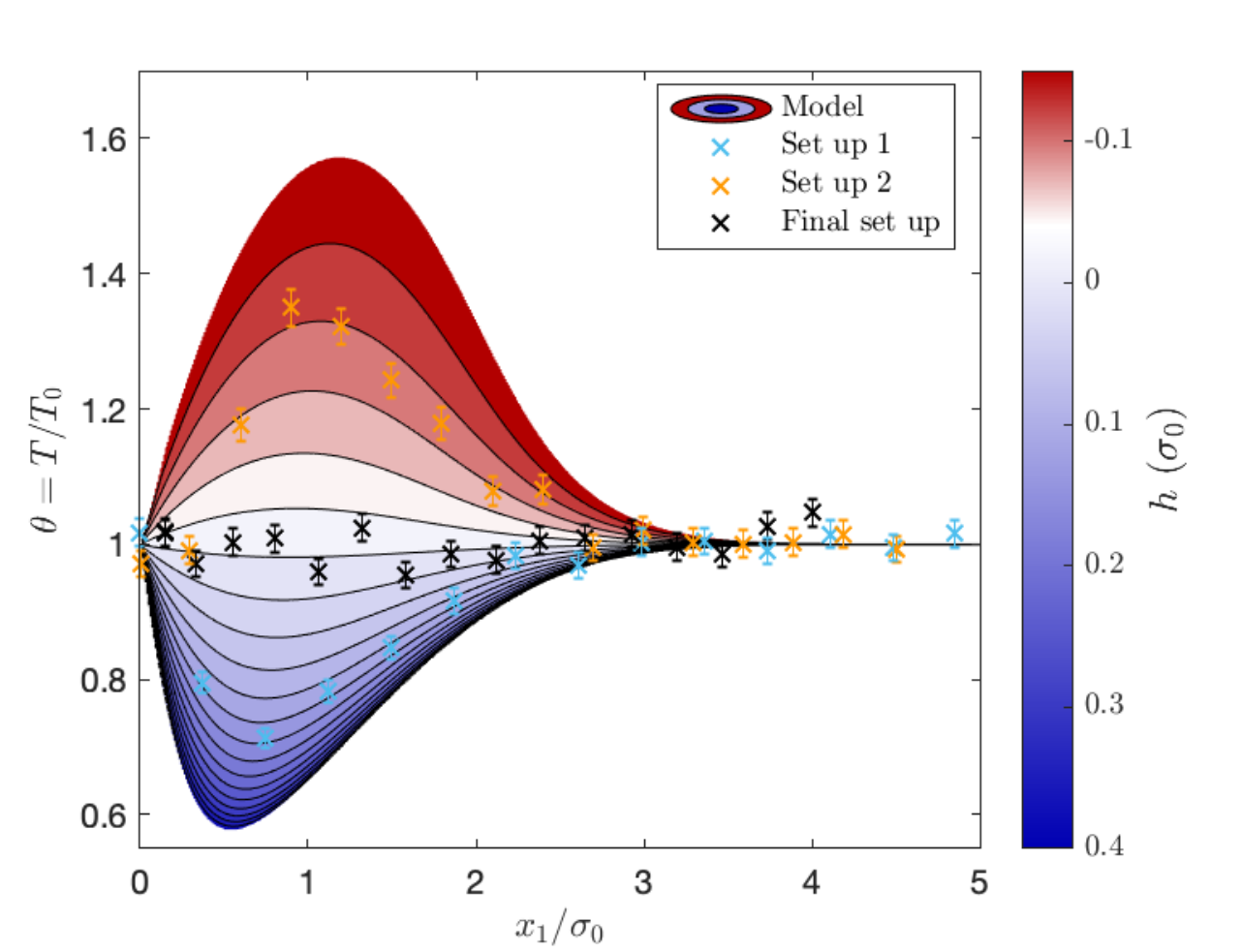}
 \end{center}
 \caption{\textbf{Consequences of an hysteresis on the kinetic temperature.} $\theta=T/ T_0$ is plotted as a function of the distance $x_1$ between the wells. The colormap is drawn with the model prediction provided by Eq.~\eqref{eqhyst}: positive hysteresis $h$ cools the system down, while negative hysteresis warms the system up. The dependance on $x_1$ comes from the balance between the barrier crossing rate and the energy step due to the hysteresis at each switch. The blue, black and orange points correspond respectively to the experimental results obtained with the three setups addressed in this paper: setup 1 ($h=0.15 \sigma_0$), final setup (tiny hysteresis), and setup 2 (early trigger).}
 \label{theta_map} 
\end{figure} 

The temperature solution of Eq.~\eqref{eqhyst} allows the system to reach a steady state in which the average heat flux lost by the system ($\Gamma \Delta U_h$), and the heat influx from the heat bath (proportional to $T-T_0$) equilibrate. The numerical solution of Eq.~\eqref{eqhyst} is plotted on Fig.~\ref{theta_map}: the kinetic temperature presents a minimum around $x_1\lesssim \sigma_0$, which deepens as the hysteresis $h$ increases. These trends can be easily understood: firstly, the larger the hysteresis, the greater the energy loss at each switch, and, therefore, the lower the system temperature. Secondly, the energy loss per switch is proportional to $x_1$, but the barrier crossing rate decreases with $x_1$: in the high barrier limit there are no more switches and $T=T_0$, and in the low barrier limit there is no more energy step at the switch, so that $T=T_0$ as well. The effect on the temperature is maximized for $x_1\sim\sigma_0$ when the two opposing effects counteract each other most. The model, applied to the setup 1 measured hysteresis, is in perfect agreement with the experimental data as highlighted in Fig.~\ref{theta_2_models}. Let us also point out that the same description holds for negative hysteresis: early switches make the system warm up, as shown in red on Fig.~\ref{theta_map}. In conclusion, removing all hysteresis at the barrier crossing is mandatory to maintain a proper equilibrium in the double-well potential, instead of creating an out-of-equilibrium steady state characterized by a temperature $T\neq T_0$.
 
\subsubsection{Simulation confirmation}
We complete the study by simulating $N_{\textrm{sim}}=200$ trajectories of the cantilever evolving in a potential created with $h=0.15 \sigma_0$ hysteresis on the threshold. The numerical simulation is in very good agreement with both the model and the experimental data (see Fig.~\ref{theta_2_models}).

\subsection{Switching delay}
A time delay between the cantilever crossing the barrier and the force switching is inevitable because real comparators have finite switching speed, but also due to the delay inherent to the low-pass filter applied to the position measurement. The effect of such a delay is similar to that of an hysteresis. Indeed, if there is a time delay $\tau_d$, the cantilever overreaches the barrier of a distance $h_d$ on average at each passage, that can be computed knowing the speed PDF: 

\begin{align}
 h_d& =\langle |v|\rangle\tau_d = \int_0^{\infty}  |v| \frac{e^{-\frac{v^2}{2 \sigma_v^2}}}{\sigma_v \sqrt{2\pi}}dv\tau_d  \\
 &=\sqrt{\frac{2\theta}{\pi}}\sigma_0 \omega_0 \tau_d 
 \end{align}
The absolute value in the average of $v$ comes from the fact that only the velocity sign that matches the barrier crossing is considered (for example positive velocity for upward crossing). 
The time delay can thus be treated as a mean hysteresis $h_d $, associated to an energy step $\Delta U_d =2kx_1h_d$, leading to an equation equivalent to Eq.~\eqref{eqhyst} with an updated barrier height:
\begin{align}
\Gamma(\frac{1}{2}kx_1^2,\theta T_0)\times \Delta U_d &=\frac{\omega_0}{Q}k_B T_0 (1-\theta) \label{eqdelay}
\end{align}
Thus, the temperature of the system trapped in a double-well potential with switching time delay $\tau_d$ is a solution of the following equation, derived from Eq.~\eqref{eqdelay}:
\begin{align}
g\left(\frac{x_1}{\sigma_0\sqrt{2\theta}}\right) Q \omega_0 \tau_d \, \theta =1-\theta \label{eqTdelay}
\end{align}
where
\begin{equation}
g(z) = \frac{4}{\sqrt{\pi}} z^{3} \int_1^\infty \frac{\exp(-\epsilon z^2)}{\pi + 2 \sin^{-1}(\epsilon^{-1/2})}  d\epsilon
\end{equation}
The numerical solution of Eq.~\eqref{eqTdelay} has a profile similar to the solutions of Eq.~\eqref{eqhyst} plotted on Fig.~\ref{theta_map}.

The function $g(z)$ presents a global maximum $g^*=0.21$ in $z^*=0.64$, allowing to compute the minimum temperature and corresponding well distance
\begin{align}
\theta_\mathrm{min} &= \frac{1}{1+g^*Q\omega_0\tau_d} =  \frac{1}{1+g^*\tau_d/\tau_\gamma} \label{eq:thetamin}\\
x_{1,\mathrm{min}} &= z^* \sigma_0\sqrt{2\theta_\mathrm{min}}
\end{align}
The minimum temperature is thus a function of the ratio between the switch delay $\tau_d$ and the smallest intrinsic time of the resonator, $\tau_\gamma$: no kinetic temperature change is expected if the former is much smaller than the latter.

\subsection{Measurement noise}
\label{noise}
The PSD in Fig.~\ref{FigPSD} demonstrates that, in a single well, the thermal noise of the cantilever is very close to that of an ideal SHO, on a wide frequency range. Nevertheless, 2 sources of deviation can be noticed. First, higher-order deflection modes (from the third up) are clearly visible, and contribute to the measured signal by adding high frequency noise accounting for $0.05\sigma_0$. Second, some background noise remains, due to higher conditioning electronic noise and to the shot noise of the photodiodes of the interferometer. At high frequencies, this noise floor, around $\SI{3e-9}{\sigma_0^2/Hz}$, supersedes the signal from the first deflection mode. Integrated on the $\SI{1}{MHz}$ bandwidth of the final setup filtering (detailed later in section \ref{device}), this background noise contributes up to $0.05 \sigma_0$. This measurement noise reaching in total $0.07\sigma_0$ has two unwanted consequences on the feedback generated potential: parasitic switches and early triggering.

\subsubsection{Parasitic swiches}

If the apparatus compares the raw deflection signal $V_{x}$ from the interferometer directly to the threshold $V_{x_0}$, the noise in the input signals produces multiple transitions at the crossing. As a consequence, the feedback loop output voltage $V$ oscillates rapidly between positive and negative values, so the mean voltage seen by the electrode vanishes. Because of these parasitic switches of the comparator, the cantilever ends up trapped at the threshold position $x_0=0$, in between the two desired equilibrium ones $\pm x_1$. 

 \subsubsection{Early triggering}
 
The high frequency noise triggers the switch before the signal of interest (the position of the first deflection mode) actually crosses the threshold, and therefore induces early switches. In setup 2 (whose circuit is detailed in Fig.~\ref{electric-setup}, and summarized into Tab.~\ref{Tab_setup}), the high frequency noise is not removed, so that a negative hysteresis appears due to the early triggers. Consequently the system temperature rises in accordance with the prediction of previous sections. The experimental evidence of the temperature rise in setup 2 is superimposed with orange crosses to the theoretical curves in Fig.~\ref{theta_map}.

 \section{Requirements} \label{sec:requirements}
 
To mitigate the consequences of the experimental non-idealities listed above, we need to adapt the experimental setup. We detail in this section the essential experimental constraints to create a proper virtual potential.
 
 \subsection{Limiting the hysteresis}
\label{limit_hysteresis}
To maintain the velocity equilibrium distribution in the virtual potential, and to limit the cooling to $5\%$, from Fig.~\ref{theta_map} we deduce that the hysteresis has to be lower than $0.02 \sigma_0$. Note that this value, deduced from the model summarized in Eq.~\eqref{eqhyst}, is computed for a quality factor of $10$, and higher values of $Q$ would result in an even more stringent requirement. As regards the cooling effect, one would wish to suppress the hysteresis altogether, but a tiny hysteresis is nevertheless needed for stability purposes: the output of the comparator circuit is unstable if no reference to the input is introduced. All in all, the hysteresis of the comparator should remain between $0.5\%$ and $2\%$ of $\sigma_0$.

 \subsection{Removing parasitic switches: temporal lock-up}
 
The common workaround to the issue of repeated fast-switches is to introduce an hysteresis through a positive feedback of the output on the comparator threshold. To be effective, this strategy requires an hysteresis wider than the measurement noise, hence larger than $0.07\sigma_0$ (see section \ref{noise}). Such a large hysteresis is prohibitive in our case because of the cooling effect. As an alternative, we implement a temporal lock-up to freeze the comparator state after a switch, for roughly $1/4$ of the oscillator's natural period $1/f_0$. By the time the comparator is active again, the cantilever has evolved in the new well --on average-- long enough to reach the bottom of the well, and is therefore far enough from the threshold that an undue noise-induced switch is improbable. One drawback is that short excursions in the other well are forbidden as well. However, these events --indeed present in a real double-well potential-- are unlikely enough that removing them has no noticeable effect of the statistical properties of the virtual potential.
 
 \subsection{Removing early triggering: low-pass filtering}
  
To correct early switches (occurring in setup 2), we must filter the high-frequency noise. The second mode contribution is hidden by focusing the laser on the mode vibration node. The higher-order modes and the electronic shot noise are low-pass filtered. When designing this filter, the concern is the delay introduced, since it will induce an hysteresis, possibly cooling the system.

On the one hand the filter has to cut the high frequency noise over $1000 f_0$ to limit the background noise contribution (increasing at high frequencies) to $0.05 \sigma_0$. But on the other hand the filter response time $\tau_d$ has to remain much lower than $\tau_\gamma/g^*$ to limit the cooling effect: this bound corresponds to $1-\theta \ll 1$ using Eq.\eqref{eq:thetamin}. To summarize, using the relation between the cutoff frequency $f_c$ of a first order low-pass filter and its response time $\tau_d\sim\frac{5}{2\pi f_c}$ (the relation holds for higher-order filters in first approximation), $f_c$ is bounded by:
\begin{align}
 \label{Eqfcrange}
5 g^* Q f_0 &\ll f_c < 1000 f_0
\end{align}

With a quality factor $Q=10$ and a resonance frequency $f_0=\SI{1.2}{kHz}$, the interval reads: $ \SI{13}{kHz}\ll f_c < \SI{1.2}{MHz}$. 
   
 \subsection{Characteristics of the cantilever}
 
The cantilever is chosen to meet the requirements of the filter cutoff frequency and the comparator hysteresis, expressed in Eq.\eqref{Eqfcrange}: by selecting a low $Q$ and low $f_0$, we minimize the cooling, and alleviate the constraints on the feedback characteristics. We thus choose $Q=10$ and a relatively slow oscillator: $f_0=\SI{1.2}{kHz}$. Furthermore, we choose a low stiffness $k\sim\SI{5e-3}{N/m}$ to have a large thermal noise, thus a large signal (Brownian) to noise (background) ratio.

\section{Final setup}  \label{sec:finalsetup}
 
   \begin{figure}[!htb]
 \begin{center}
 \includegraphics[width=\columnwidth]{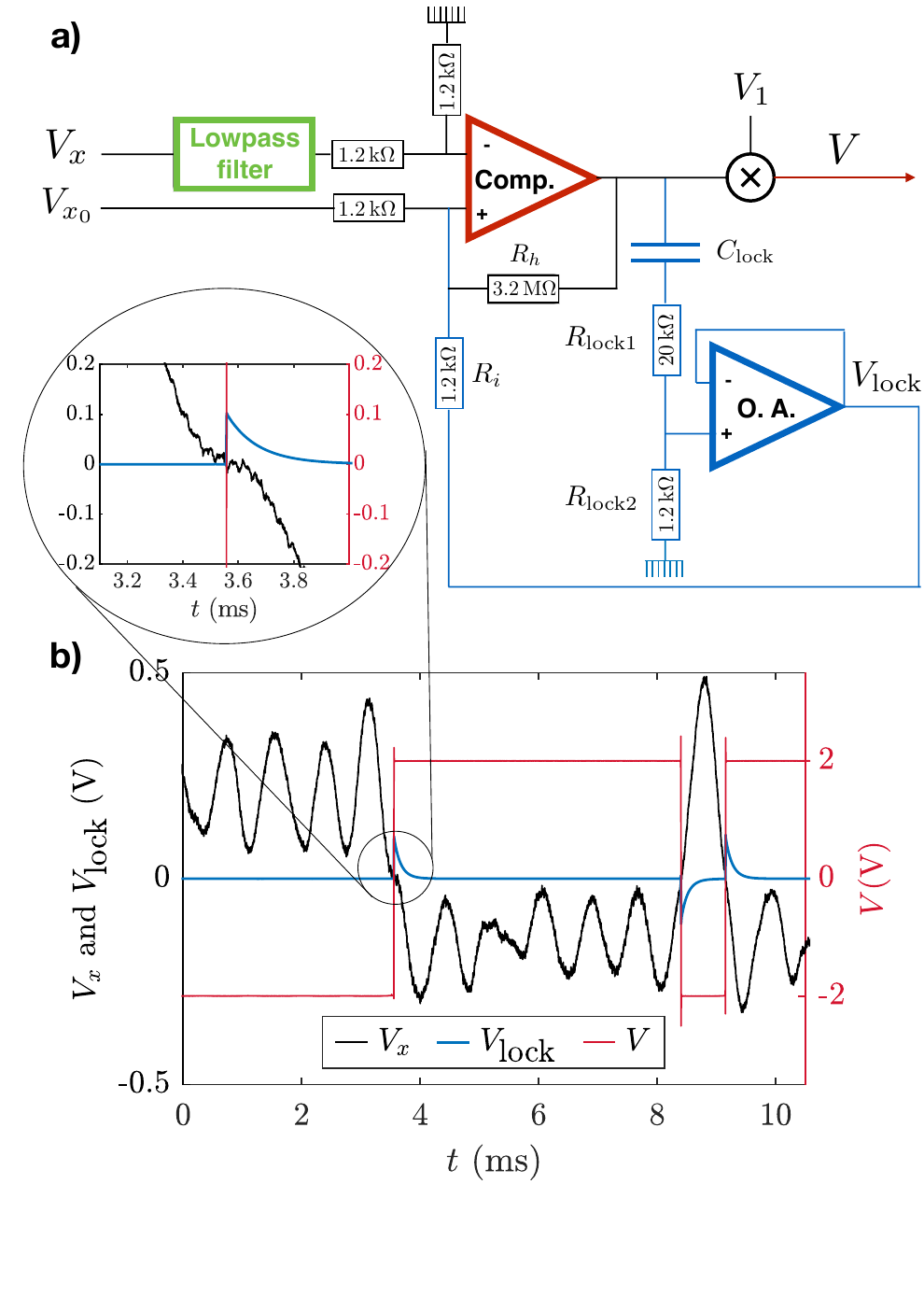}
 \end{center}
 \caption{\textbf{(a) Electrical diagram of the feedback loop.} The cantilever deflection signal $V_x$ from the interferometer passes through a low pass filter ($f_c=\SI{1}{MHz}$, model SR560, green) before entering the LM219 comparator (red). The threshold signal $V_{x_0}$ is momentarily modified by $V_{\textrm{lock}}$ after each switching of the comparator. $V_{\textrm{lock}}$ is the result of a lock-up feedback consisting in a follower assembly and a capacitive circuit built with the following components: operational amplifier LT131, capacity $C_{\textrm{lock}}=\SI{4.7}{nF}$, resistances $R_{\textrm{lock1}}=\SI{20}{k\Omega}$ and $R_{\textrm{lock2}}=\SI{1.2}{k\Omega}$. The output of the comparator is then multiplied by $V_1$ to modulate the final voltage $V$ (using an AD633-EVALZ analog multiplier). With respect to this design (denoted as final setup), we call setup 1 the same circuit but with a TS3022 comparator leading to a switching hysteresis $h=0.15\sigma_0$. Similarly we call setup 2 the final circuit without the low-pass filter leading to early triggers (negative hysteresis). The setups' distinctive characteristics are summarized in table~\ref{Tab_setup}. \textbf{(b) Example of signals.} The cantilever deflection signal $V_x$ is plotted in black, the lock-up voltage $V_{\textrm{lock}}$ in blue and the output voltage $V$ in red (with $V_1=2V$). The threshold $V_{x_0}$ is set to $0$.}
  \label{electric-setup}
\end{figure}

We detail in this section the final experimental setup designed to meet all the requirements previously listed. The feedback circuit diagram is detailed in Fig.~\ref{electric-setup}: it contains the basic components (comparator and multiplier) on which some elements are added to ensure its efficiency. The deflection signal from the interferometer $V_{x}$ is filtered by a low-pass filter (green) before entering the comparator device (red). The tunable threshold $V_{x_0}$ is biased by the voltage $V_{\textrm{lock}}$ resulting from the lock-up feedback loop (blue components) before being compared to $V_{x}$. The comparator output voltage is then multiplied by the adjustable voltage $V_1$.

\subsection{Stability}

To ensure the stability of the comparator output, we introduce a tiny hysteresis corresponding to $R_i/R_h=\SI{1.2}{k\Omega}/\SI{3.2}{M\Omega}$ fraction of the output fed on the positive input (see Fig.~\ref{electric-setup}). As the position signal scales as $\sigma_0  \textrm{ (in V)}=\sqrt{\langle V_x^2 \rangle} \sim \SI{50}{mV}$, the hysteresis has to stay between $0.5\% \sigma_0= \SI{0.25}{mV}$ and $2\% \sigma_0=\SI{1}{mV}$ to meet the requirements of section \ref{limit_hysteresis}. With the $\pm\SI{1}{V}$ power supply voltage of the comparator device, the hysteresis amplitude of our final design reaches $\SI{0.37}{mV}$ and therefore remains in the range specified.

\subsection{Temporal lock up}

The temporal lock-up feedback is implemented through a follower assembly and a capacitor (blue components on Fig.~\ref{electric-setup}). The comparator (red device on Fig.~\ref{electric-setup}) compares $V_x/2$ to $(V_{x_0}+V_{\textrm{lock}})/2$, without being affected by the temporal lock-up components values thanks to the impedance conversion provided by the buffer. The purpose of this is to bias the threshold $V_{x_0}$ during the discharge time of the capacitor $C_{\textrm{lock}}$, in order to prevent the comparator switching back right after a switch. In the static regime without switches, the output is constant for example at $+V_{\textrm{sat}}=\SI{1}{V}$, which corresponds to the charged capacity that acts as an open circuit so that $V_{lock}=0$. Right after a switch of the output voltage, the capacity starts reversing its charge through $R_{\textrm{lock2}} +R_{\textrm{lock1}}$, and $V_{\textrm{lock}}$ moves immediately to $2R_{\textrm{lock2}}/ (R_{\textrm{lock2}} +R_{\textrm{lock1}}) V_{\textrm{sat}}=\SI{110}{mV}$, before decreasing to $0$. As long as $V_{lock}$ remains large, it prevents any switch. The capacity $C_{\textrm{lock}}=\SI{4.7}{nF}$ rules the $V_{\textrm{lock}}$ relaxation time $\tau_{\textrm{lock}}=(R_{\textrm{lock1}}+R_{\textrm{lock2}}) C_{\textrm{lock}}=\SI{0.1}{m s}$. It is chosen to freeze the comparator during approximately a quarter of the cantilever period: $3 \times \tau_{\textrm{lock}}=\SI{0.3}{ms}\sim1/(4 f_0)$. Indeed, we verify on Fig.~\ref{electric-setup} b) that when $V_{x}$ (black line) crosses the threshold $V_{x_0}=0$, the comparator properly switches only once from $V=-V_1$ to $V=+V_1$ (red line), as $V_{\textrm{lock}}$ (blue line) becoming transiently positive significantly increases the threshold value for approximately $\SI{0.3}{ms}$. 

\subsection{Devices characteristics}
\label{device}
To maintain less than $5\%$ cooling in the final setup, we use a LM219 comparator that has no hysteresis and a typical $\SI{80}{ns}$ response in the working conditions. A tiny hysteresis of $\SI{0.37}{mV}$ is added through feedback resistances to guarantee stability: $R_i/R_h=\SI{1.2}{k\Omega}/\SI{3.2}{M\Omega}$. 
Finally, the low-pass filter added to remove early triggers has its cutoff frequency chosen within the prescribed range: $f_c=\SI{1}{MHz}$. A smaller cutoff frequency could be chosen (down to $\sim \SI{500}{kHz}$ as prescribed by Eq.\eqref{Eqfcrange}), to lower the background noise contribution even more (reduced to $4\% \sigma_0$ for a $\SI{500}{kHz}$ bandwidth). In the final setup, we use a SR560 model containing two identical tunable cutoff frequencies 1st-order R-C filters, to provide first or second-order filtering at $f_c=\SI{1}{MHz}$. 

\subsection{Virtual potential characteristics}

The position distribution of the cantilever trapped in the virtual potential produced with the final setup perfectly matches the expected equilibrium distribution in a double-well, as illustrated in Fig.~\ref{FigDoubleWell}. Moreover, we show in Fig.~\ref{FigPDFspeed} that the velocity distribution in closely-spaced wells ($x_1=\sigma_0$) is also in excellent agreement with the equilibrium expectation (without feedback), contrary to the two previous setups, for which the velocity PDF is clearly modified by the feedback. To complete the experimental verification, we measure the velocity variance for different distances between the wells. Fig.~\ref{theta_map} (black markers) shows that the velocity distribution in the virtual double-well potential of the final setup is not biased.

\section{Conclusion}

\begin{figure}[!htb]
 \begin{center}
 \includegraphics[width=\columnwidth]{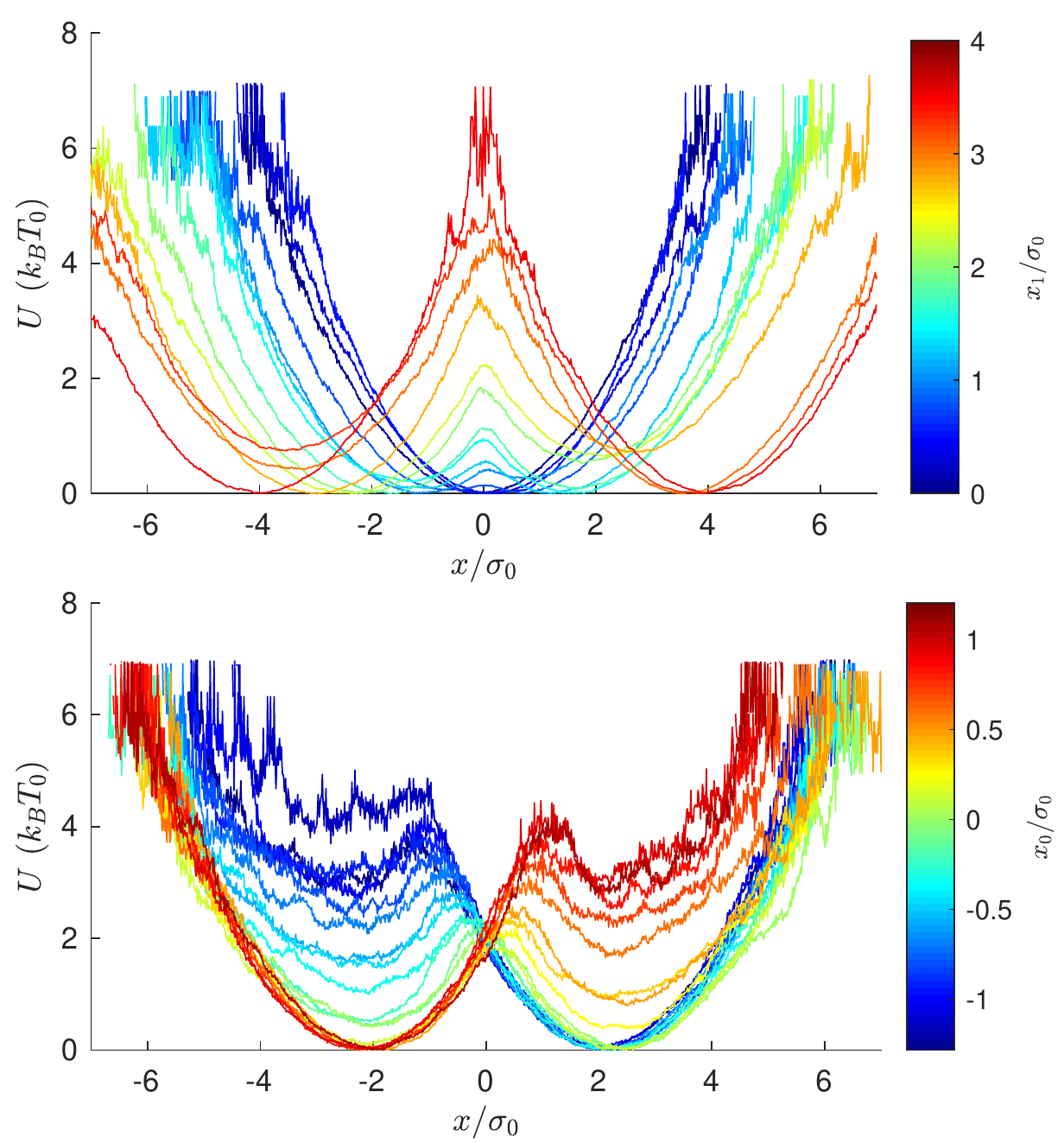}
 \end{center}
 \caption{\textbf{Double-well potential tuning}. $U(x_0,x_1,x)$ is computed through the measured PDF of $x$ during $\SI{10}{s}$ acquisitions and the equilibrium Boltzmann distribution for different values of the controlled parameters $x_0$ and $x_1$. The top graph corresponds to $x_0=0$ and $x_1\in[0,4]\sigma_0$, the bottom one to $x_0\in[-1.25, 1.25]\sigma_0$ and $x_1=2\sigma_0$. The two parameters allow to explore different barrier height and potential energy step between the two wells. The high noise level for large values of $U$ stems from the low sampling of regions with low probability.}
 \label{U} 
\end{figure} 

This underdamped system has the strong merits of a short relaxation time and a highly precise deflection measurement, but controlling its virtual potential requires special caution on the feedback control. Namely, the underdamped regime makes the response much more sensitive to any noise or delay in the driving force. The thorough study of the effects of experimental non-idealities enables us to identify the key requirements needed to create a proper virtual potential. The experimental challenge that ensues is successfully addressed by the final setup. Ultimately, we demonstrate that the response of the system in the double-well potential built this way is statistically equivalent to the one expected at equilibrium in a physical potential.
 
Additionally, this virtual potential can be precisely controlled through the tuneable parameter $x_0$ which sets the barrier position, and the parameter $x_1$ which defines the distance between the wells. Fig.~\ref{U} shows the influence of the above mentioned parameters on the potential shape. The wells curvature is not tuneable as solely set by the cantilever stiffness, and the barrier height is enslaved to all other parameters.

The experimental work detailed in this article opens a wide range of possibilities in the field of underdamped system control, and allows high accuracy exploration of statistical physics in the underdamped regime (and in particular stochastic thermodynamics). The analysis presented holds at even lower damping $Q\gg1$, achievable by placing the cantilever in vacuum. This configuration simply imposes more stringent constraints on the feedback time delay. Finally, this electrical circuit paves the way to the use of a field-programmable gate array (FPGA) configured to perform all the calibration and feedback operations, improving reliability and accuracy. Indeed, such a digital controller can readily give microsecond response (or even faster) and would meet easily the experimental requirements listed here, even in the highly underdamped regime. Besides, more complex configurations of the FPGA target (associating a specific output voltage to every position) would even allow to create any arbitrary non-linear potential shape, with several applications from optimal protocols for minimizing the work in finite-time operations~\cite{Gomez, Aurell_2012}, to the exploration of non-equilibrium extensions of Landauer's theory~\cite{Sagawa_2014,Esposito_2011}.

\acknowledgments 
\noindent \textbf{Acknowledgments} This work has been financially supported by the Agence Nationale de la Recherche through grant ANR-18-CE30-0013 and by the FQXi Foundation, Grant No. FQXi-IAF19-05, “Information as a fuel in colloids and super-conducting quantum circuits.”.

\medskip
\noindent \textbf{Data availability} The data that support the findings of this study are openly available in Zenodo~\cite{Dago-2022-JStat}. 

\appendix

\section{Mean Heat} \label{AppendixB}
 We derive in this section the very general expression of the average heat over an underdamped stochastic process following Ref.~\citenum{sek10}.

Applying to the underdamped regime the generic computations of stochastic energy exchanges~\cite{sek10,DagoPRL,Aurell_2012,Seifert_2012,Jarzynski_2011}, we have: 
\begin{align}
\frac{d\Q}{dt}&=-\frac{\partial U}{\partial x} \dot{x} -\frac{dK}{dt}. \label{dQsdt}
\end{align}

The computation of the mean dissipated heat requires writing the general Langevin equation of an underdamped system in a potential $U$:
\begin{equation}
m\ddot x = -\frac{\partial U}{\partial x} -\gamma \dot x  +F_{th}, \label{langevingen}
\end{equation}
where $F_{th}$ is a delta correlated white Gaussian noise corresponding to the forcing due to the thermal bath: $\langle F_{th}(t)F_{th}(t+t') \rangle = 2 k_B T_0 \gamma \delta(t')$. Multiplying Eq.\eqref{langevingen} by $\dot x$ leads to the dissipated heat defined by Eq.~\eqref{dQsdt}:
\begin{align}
\frac{d\Q}{dt}=m\ddot x \dot x -\frac{dK}{dt} + \gamma \dot x ^2  -F_{th}\dot x. \label{computeQ}
\end{align}

Some caution is required before taking the mean value of the above expression, because it involves products of stochastic quantities: in that respect, the Ito discretization prescribes for a stochastic function $K(v)$, 
\begin{align}
\frac{dK}{dt}=\frac{\partial K}{\partial v} \dot v +\frac{1}{2} \frac{\partial^2 K}{\partial v^2}\dot v^2 dt. \label{Ito}
\end{align}
We apply Eq.~\eqref{Ito} to $K=\frac{1}{2} m v^2$, and use Eq.~\eqref{langevingen} to compute the $\dot v^2$ term:
\begin{align}
\frac{dK}{dt}=m v \dot v +\frac{1}{2m} \left(  -\frac{\partial U}{\partial x} -\gamma \dot x  +F_{th} \right)^2 dt
\end{align}
When taking the mean value and letting $dt$ tend to $0$, most terms simplify out. Indeed, only remain the terms that involve the thermal noise $F_{th}$ scaling in $1/\sqrt{dt}$, some of which are cancelled by the Ito prescription: $\langle F_{th}  v \rangle = \langle F_{th}  x \rangle=0$.  
Finally, we obtain the relation: $d \langle K \rangle/dt = m\langle \ddot x \dot x \rangle + k_B T_0 \gamma/m$. Eq.~\eqref{computeQ} then simplifies into:
\begin{align}
\frac{d\langle \Q\rangle}{dt}&=\frac{\gamma}{m}(2  \langle K \rangle - k_B T_0). \label{Qgen}
\end{align}
Using the definition of the kinetic temperature $T=2 \langle K \rangle /k_B$, and introducing the quality factor $Q=m\omega_0/\gamma$, Eq.~\eqref{Qgen} becomes: 
\begin{align}
\frac{d\langle \Q\rangle}{dt}&=\frac{\omega_0}{Q}k_B (T - T_0). \label{Qstat}
\end{align}
This expression is completely general and highlights that the heat exchanges are reduced at high $Q$~\cite{dago2021fast}.

\section{Switching rate} \label{AppendixA}

In the limit of weak damping, the total energy of the cantilever $E=U+K$ is conserved, and its motion is periodic in time. The period of oscillation $\mathcal{T}$ depends on the value of $E$ with respect to the barrier height $\mathcal{B}=\frac{1}{2} k x_1^2$. If $E<\mathcal{B}$, then the motion is confined to a single well, there is no switches, and the period is $\mathcal{T}_0=1/f_0$. If $E>\mathcal{B}$, the cantilever visits both wells every period, so there are 2 switches every period, with
\begin{align}
\mathcal{T}(E,\mathcal{B}) &= 2 \int_{-x_M}^{x_M} \sqrt{\frac{m}{2(E-U(x,x_1))}} dx \\
&= \frac{2}{\omega_0} \int_{-x_M}^{x_M} \frac{1}{\sqrt{(x_M-x_1)^2-(|x|-x_1)^2}} dx \\
&= \frac{2}{\omega_0} \left[\pi+ 2 \sin^{-1}\left(\sqrt{\frac{ \mathcal{B}}{E}}\right)\right],
\end{align}
where $x_M = x_1 +\sqrt{2E/k}$ is the maximum excursion of the cantilever. This period is twice $\mathcal{T}_0$ when $E\gtrsim\mathcal{B}$, and tends to $\mathcal{T}_0$ for $E\gg\mathcal{B}$.

In equilibrium, the statistics of the total energy $E$ is ruled by the Boltzmann distribution: $P(E)= \exp(-E/k_B T)/k_B T$. We deduce the average switching rate $\Gamma$ by weighting the 2 switches per period for $E>\mathcal{B}$ by this probability:
\begin{equation} \label{EqGamma}
\Gamma(\mathcal{B},T) = \int_\mathcal{B}^\infty \frac{2}{\mathcal{T}(E,\mathcal{B})} \frac{\exp(-E/k_B T)}{k_B T} dE.
\end{equation}

For finite damping, the energy is not conserved along single trajectories, but still is in average thanks to the equilibrium with the thermostat. Eq.~\eqref{EqGamma} is therefore a good approximation of the switching rate between the wells for a given barrier height and system temperature.

\begin{figure}[!bht]
\begin{center}
\includegraphics[width=\columnwidth]{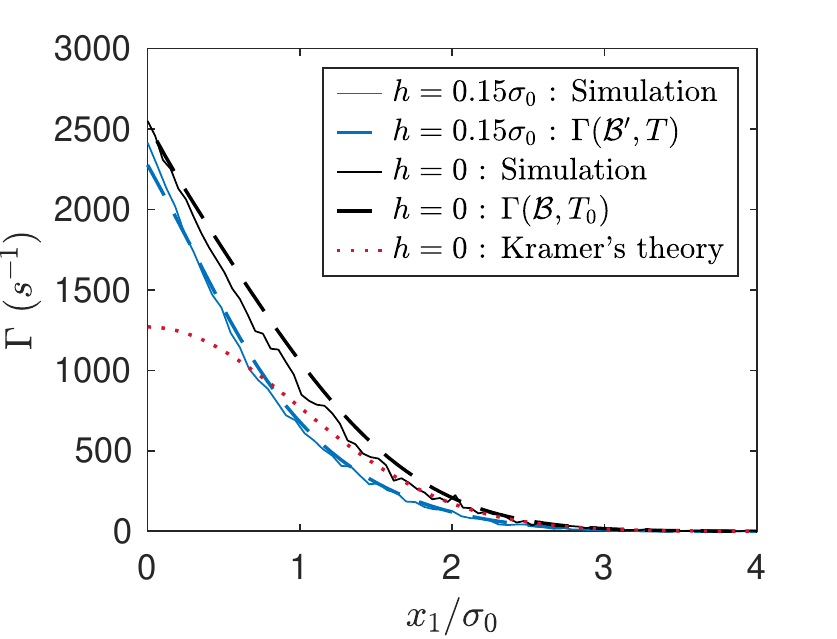}
\end{center}
\caption{\textbf{switching rate} $\Gamma$ as a function of the distance $x_1$ between the wells without switching delay, for two hysteresis: $h=0$ and $h=0.15 \sigma_0$. For $h=0$ and high energy barriers $\mathcal{B}=\frac{1}{2}x_1^2$, Kramer's simplest model in dashed red line holds and perfectly matches the simulation data from $N=100$ iterations of $30/f_0$ long trajectories in black line. However, for smaller barrier height the model $\Gamma(\mathcal{B},T_0)$ of Eq.~\eqref{EqGamma} in black dashed line provides a better prediction. For a $h=0.15 \sigma_0$ hysteresis, the simulation data from $N=100$ iterations of $30/f_0$ long trajectories in blue line is in very good agreement with the model $\Gamma(\mathcal{B}',T)$ with parameters $T=\theta_h T_0$ solution of Eq.~\eqref{eqhyst} and $\mathcal{B}'=\frac{1}{2}(x_1+h)^2$.}
 \label{Gamma} 
\end{figure}

In Fig.~\ref{Gamma}, we superimpose the switching rate computed with Eq.~\eqref{EqGamma} and the results of a simulation with or without hysteresis. The good agreement between the simulation and the model justifies the use of the $\Gamma(\mathcal{B},T)$ expression to derive the temperature evolution of the system in a double-well potential with switching delay.
Besides, the dotted red line corresponds to Kramer's theory~\cite{KRAMERS} prescribing the escape rate $\Gamma_K(\mathcal{B},T)=\frac{\omega_0}{2 \pi} e^{-\mathcal{B}/k_BT}$~\cite{melnikov_kramers_1991}. Hence, Fig.~\ref{Gamma} highlights the fact that Kramer's simplest formula $\Gamma_K$ doesn't work for low energy barriers.

\bibliography{UDfeedbackpotential}

\end{document}